\documentstyle[a4]{article}
\begin{document}

\newcommand{\gtrsim}{
\,\raisebox{0.35ex}{$>$}
\hspace{-1.7ex}\raisebox{-0.65ex}{$\sim$}\,
}

\newcommand{\lesssim}{
\,\raisebox{0.35ex}{$<$}
\hspace{-1.7ex}\raisebox{-0.65ex}{$\sim$}\,
}

\newcommand{\const}{{\rm const}}

\begin{flushleft}
{\small J. Phys. A: Math. Gen. {\bf 29} (1996) 
{\footnotesize L}257--{\footnotesize L}262. Printed in Hamburg}
\vspace{0.7cm}
\end{flushleft}

\begin{flushright}
\parbox[t]{11cm}{
{\large\bf LETTER TO THE EDITOR}
} 
\end{flushright}

\vspace{0.5cm}

\begin{flushleft}
{
\Large\bf
Spherical model for anisotropic ferromagnetic films
}      
\end{flushleft}

\vspace{0.1cm}

\begin{flushright}
\parbox[t]{11cm}{

D A Garanin 
\renewcommand{\thefootnote}{\fnsymbol{footnote}}
\footnotemark[2]
\vspace{0.1cm}

\small
I. Institut f\"ur Theoretische Physik, Universit\"at Hamburg,
Jungiusstr. 9, \\ 
D-20355 Hamburg, Germany\\
\vspace{0.2cm}

Received 26 October 1995
\vspace{0.4cm}

{\bf Abstract.} \hspace{1mm}
The corrections to the Curie temperature $T_c$ of a ferromagnetic film 
consisting of $N$ layers are calculated for $N\gg 1$ for the model of 
$D$-component classical spin vectors in the limit $D\to\infty$, which is 
exactly soluble and close to the spherical model. The present approach 
accounts, however, for the magnetic anisotropy playing the crucial role 
in the crossover from 3 to 2 dimensions in magnetic films. In the 
spatially inhomogeneous case with free boundary conditions the 
$D=\infty$ model is nonequivalent to the standard spherical one 
and always leads to the diminishing of $T_c(N)$ relative to the bulk. 
} 

\vspace{1.2cm}

\end{flushright}

\renewcommand{\thefootnote}{\fnsymbol{footnote}}
\renewcommand{\footnoterule}{\rule{0cm}{0cm}}
\footnotetext[2]{E-mail: garanin@physnet.uni-hamburg.de}

The application of the spherical model \cite{berkac52} to 
spatially-inhomogeneous magnetic systems such as 
ferromagnetic films with free boundary conditions 
by Barber and Fisher \cite{barfis73} has revealed the unphysical 
behaviour of the solution being the consequence of the global spin 
constraint. The dependence $T_c(N)$ 
for a $d$-dimensional hypercubic lattice infinite in $d'=d-1$ dimensions 
and having $N$ layers in the $d$th dimension has been found 
to be for $d\geq4$ a non-monotonous function with a maximum, i.e., 
$T_c(N)$ for $N\gg 1$ was larger than in the bulk. Other singular 
features of the spherical model were found by Abraham and Robert 
\cite{abrrob80} by considering the problem of 
phase separation (i.e., the domain wall formation).

Besides the numerous publications using the spherical model 
for inhomogeneous systems in its original form
(see, e.g., \cite{allpat93,pat94}), there is 
a work by Costache, Mazilu und Mihalache \cite{cosmazmih76} in which the 
global spin constraint was replaced 
for a ferromagnetic film by separate constraints in each 
layer. Although this model is less convenient for analytical 
calculations, it was shown that for $d\geq4$ the value of 
$T_c(N)$ monotonically increases to its bulk value $T_c(\infty)$, as it 
should from the physical grounds. Earlier Knops \cite{kno73} had 
proved that in a general inhomogeneous situation the spherical model 
with a spin constraint on each lattice site is equivalent to the 
$D$-component classical vector model by Stanley \cite{sta68} in the 
limit $D\to\infty$. The latter is not only more physically appealing 
than the original spherical model, but it also allows one to take into 
account 
the spin anisotropy \cite{oka70} and to produce the $1/D$ expansions
\cite{abe7273,abehik7377,gar94jsp}. A convenient tool to 
handle the $D$-vector model is the classical spin diagram technique   
\cite{gar94jsp,garlut84d}.

The possibility of considering anisotropic systems makes the 
analytically soluble $D=\infty$ model, which can also be called for 
simplicity the spherical model, rather attractive for applications. 
In Ref.\cite{gar96jpa}
it was used to investigate the role of fluctuations in 
the phase transition from Bloch to linear domain walls in {\em biaxial} 
ferromagnets. Anisotropy also plays a 
crucial role for ferromagnetic films in the actual case $d=3$. 
For $N\ne\infty$ the system is infinite in only $d'=2$ dimensions, and 
cannot sustain a long-range order in the case of a contunuous spin symmetry.
Correspondingly, Barber and Fisher \cite{barfis73} have found
diverging corrections to $T_c$ for $N\gg 1$ for the standard spherical 
model. On the other hand, for a purely 2-dimensional system ($N=1$) $T_c$ 
tends to zero only logarithmically slow with vanishing anisotropy. 
One can expect that in the quasi 3-dimensional case ($N\gg 1$) the 
characteristic anisotropy required to support the long-range order 
should be extremely small. The calculation of the $T_c$-corrections for 
3-dimensional ferromagnetic films, which strongly depend on the 
anisotropy, is the main purpose of this work.
 
The hamiltonian of the anisotropic classical $D$-vector model can be 
written in the form
\begin{equation}\label{dham}
{\cal H} = 
- \frac{1}{2}\sum_{ij}J_{ij}
\left(
m_{zi}m_{zj}  
+ \eta \sum_{\alpha=2}^D m_{\alpha i} m_{\alpha j}
\right) ,
\end{equation}
where ${\bf m}_i$ is the normalized $D$-component vector, 
$|{\bf m}_i|=1$ and $\eta < 1$ is the dimensionless anisotropy factor. 
In the mean field approximation (MFA) the Curie temperature of 
this model is $T_c^{\rm MFA}=J_0/D$, where $J_0$ is the zero Fourier 
component of the exchange interaction. It is convenient to introduce the 
dimensionless temperature variable $\theta\equiv T/T_c^{\rm MFA}$ and the 
reduced correlation function (CF) of transverse ($\alpha\geq 2$) spin 
components: $s_{ij}\equiv D\langle m_{\alpha i}m_{\alpha j}\rangle$, 
which are well-behaved in the limit $D\to\infty$. Using the diagram 
technique for classical spin systems \cite{garlut84d,gar94jsp,gar96jpa}, 
one arrives in the limit $D\to\infty$ at the closed 
system of equations for the average magnetization 
$m_i\equiv \langle m_{zi}\rangle$ and the CF $s_{ij}$. 
These are the magnetization equation 
\begin{equation}\label{sphermageq}
m_i = G_i \sum_j \lambda_{ij}m_j ,
\end{equation}
the Dyson equation for the correlation function 
\begin{equation}\label{dysoncf2}
s_{ii'} 
= 
\theta G_i \delta_{ii'}
+ 
\eta G_i \sum_j \lambda_{ij}s_{ji'}
\end{equation}
and the kinematic relation playing the role of the spin constraint on 
a lattice site $i$
\begin{equation}\label{sconstr}
s_{ii} + m_i^2 = 1 .
\end{equation}
Here $\lambda_{ij}\equiv J_{ij}/J_0$ and $G_i$ is the 
one-site spin average 
\mbox{$D\langle m_{\alpha i}m_{\alpha i}\rangle/\theta$} 
renormalized by fluctuations, which should be eliminated from the equations. 

In the Ising case ($\eta=0$) the influence of the fluctuations of the
transverse spin components disappear. Since, additionally, the 
longitudinal fluctuations dying out as $1/D$ are not present in the 
equations above, the situation is in this case exactly described by the MFA. 
Equation (\ref{dysoncf2}) has for $\eta=0$ the trivial solution 
$s_{ii}=\theta G_i$; then from constraint (\ref{sconstr}) one gets 
$G_i=(1-m_i^2)/\theta$, and the elimination of $G_i$ in (\ref{sphermageq})
leads to a closed equation for the magnetization. 

In the homogeneous case $m_i=m$ and $G_i=G$ 
are constants, and the equation (\ref{dysoncf2}) can be easily 
solved with the help of the Fourier transformation, which results in
\begin{equation}\label{cfhomo}
s_{ii} = 
v_0\!\!\!\int\!\!\!\frac{d^d{\bf k}}{(2\pi)^d} s_{\bf k} 
=
\theta G P(\eta G), \qquad
P(X) \equiv 
v_0\!\!\!\int\!\!\!\frac{d^d{\bf k}}{(2\pi)^d}
\frac{1}{1-X\lambda_{\bf k}} . 
\end{equation}
Here $v_0$ is the unit cell volume and 
$\lambda_{\bf k} \equiv J_{\bf k}/J_0$. For the $d$-dimensional 
hypercubic lattices $v_0=a_0^d$, $a_0$ is the lattice spacing,   
\begin{equation}\label{lambdaka}
\lambda_{\bf k} = \frac{1}{d}\sum_{i=1}^d \cos(a_0k_i) ,
\end{equation}
and the lattice integral $P(X)$ has the following properties:
\begin{equation}\label{plims}
\renewcommand{\arraystretch}{1.2}
P(X) \cong
\left\{
\begin{array}{lll}
1 + X^2/(2d),                      & X \ll 1                           \\
(1/\pi)\ln[8/(1-X)],                & 1-X \ll 1,   & d=2 \\
W_3 - c_3(1-X)^{1/2},              & 1-X \ll 1,   & d=3 \\
W_4 - c_4(1-X)\ln[c'_4/(1-X)],     & 1-X \ll 1,   & d=4,       
\end{array}
\right. 
\end{equation}
where $W_3=1.51639$ and $W_4=1.23947$ are the Watson 
integrals, $c_3 = (2/\pi)(3/2)^{3/2}$ and $c_4 = (2/\pi)^2$. 
For $d\geq 5$ the 
leading terms of the expansion of $P(X)$ about $X=1$ are non-singular.
Since in the homogeneous case the sum in the right-hand side of
(\ref{sphermageq}) equals to $m$, it is satisfied only if 
$m=0$ (above $\theta_c$) or $G=1$ (below $\theta_c$). Then
from equation (\ref{sconstr}) one gets the temperature-dependent 
magnetization:
\begin{equation}\label{mspher}
m = (1-\theta/\theta_c)^{1/2}, \qquad 
\theta \le \theta_c \equiv 1/P(\eta) .
\end{equation}
In the isotropic case ($\eta=1$) for $d\geq 3$ the value of the transition  
temperature in the bulk $\theta_c$ reduces to the well-known result 
$\theta_c=1/W$ \cite{berkac52}. For $d=2$ one obtains 
$\theta_c(\eta)\cong \pi/\ln[8/(1-\eta)]$ 
vanishing for $\eta\to 1$. 
In the Ising case $\eta=0$ the MFA result $\theta_c=1$ is reproduced.

To solve the equations of the spherical model for a $d$-dimensional 
hypercubic ferromagnetic film it is convenient to use the Fourier 
representation in $d'=d-1$ translationally-invariant dimensions and 
the site representation in the $d$th dimension. The Dyson equation 
(\ref{dysoncf2}) for the Fourier-transformed CF $\sigma_{nn_0}({\bf k})$ 
takes on the form of a system of the second order finite-difference 
equations 
\begin{equation}\label{dysoncffd}
2b_n\sigma_n - \sigma_{n+1} - \sigma_{n-1} 
= (2d\theta/\eta)\delta_{nn_0}, \qquad n=1,2,...,N ,
\end{equation}
where the mute index $n_0$ of $\sigma$ was dropped. For the free and 
periodic boundary conditions (fbc and pbc) in (\ref{dysoncffd}) we set
\begin{equation}\label{bcond}
\begin{array}{ll}
\sigma_0 = \sigma_{N+1} = 0, \qquad     & {\rm (fbc)} \\
\sigma_0 = \sigma_N, \qquad \sigma_{N+1} = \sigma_1, 
\qquad & {\rm (pbc, \, N\geq 3)} . 
\end{array}
\end{equation}
The coefficient $b_n$ in (\ref{dysoncffd}) reads
\begin{equation}\label{bn}
b_n = 1 + d[(\eta G_n)^{-1}-1] + d'(1-\lambda'_{\bf k}) ,
\end{equation}
where $\lambda'_{\bf k}$ is given by (\ref{lambdaka}) with 
$d\Rightarrow d'$. The magnetization equation (\ref{sphermageq}) takes 
on the form 
\begin{equation}\label{mageqfd}
2\bar b_n m_n - m_{n+1} - m_{n-1} = 0 
\end{equation}
with $\bar b_n\equiv b_n(\eta\!=\!1,{\bf k}\!=\!0)$ and the boundary
conditions similar to (\ref{bcond}). The constraint 
equations (\ref{sconstr}) can now be written as
\begin{equation}\label{sconstrfd}
s_{nn} + m_n^2 = 1, \qquad 
s_{nn} = 
a_0^{d'}\!\!\!\int\!\!\!\frac{d^{d'}{\bf k}}{(2\pi)^{d'}} 
\sigma_{nn}({\bf k}) , \qquad n=1,2,...,N . 
\end{equation}
The solution of the equation (\ref{dysoncffd}) is governed by the 
effective ${\bf k}$-dependent correlation length, which in the 
long wavelength region, $a_0k\ll 1$, is given by
\begin{equation}\label{corlen}
r_c(k) = a_0/\sqrt{ 2d [(\eta G)^{-1}-1] + (a_0 k)^2 } ,
\end{equation}
and which should be compared with the film thickness $L=Na_0$. In the region 
of parameters $r_c(k) \ll L$ one can expect the $d$-dimensional 
quasi-bulk behaviour perturbed due to the finite $L$. In the opposite 
limit a behaviour corresponding to the reduced dimensionality $d'=d-1$ 
is to be expected. For $d\geq 3$ in situations where the finite-size
corrections to $\theta_c$ are small, the main contribution to the 
integral (\ref{sconstrfd}) comes from the region $a_0 k \sim 1$. For 
such wave vectors the correlation length (\ref{corlen}) is of the order 
of the lattice spacing $a_0$, and $\sigma_{nn}({\bf k})$ are the 
functions of $G_n$ in several neighbouring layers. Then from the 
constraint equations (\ref{sconstrfd}) it follows (at least in the 
paramagnetic state, $m_n=0$) that the inhomogeneity of $G_n$ in the fbc 
case is confined to the boundary regions $n, N\!-\!n \lesssim n_c \sim 1$.
Due to this inhomogeneity an analytical solution of the problem is 
possible only in limiting cases. 
In the subsequent we shall restrict ourselves to the calculation of 
the Curie temperature $\theta_c$ of ferromagnetic films with $N\gg 1$. 

In the Ising limit $\eta=0$ we have $G=1/\theta$ above $\theta_c$, and 
$\theta_c$ can be found in the fbc case from the condition 
that the determinant of the 
linear system of equations (\ref{mageqfd}) turns to zero. This
leads to the MFA result \cite{woldewhalpal71}
\begin{equation}\label{tcmfa}
\theta_c = 1 - \frac{1}{d}\left( 1 - \cos{\frac{\pi}{N+1}} \right) .
\end{equation}
For the model with pbc $G$ is also independent of $n$ 
due to the symmetry of the problem, and for $\theta\leq\theta_c$, 
where $m\ne 0$, one finds $G=1$ from (\ref{sphermageq}) or (\ref{mageqfd}). 
The homogeneous solution of the finite-difference equation 
(\ref{dysoncffd}) with $b_n=b$ has the form 
$\sigma_n = c_1 \mu^n + c_2 \mu^{-n}$, and the result for the one-layer 
correlator $\sigma_{nn}$ reads
\begin{equation}\label{sigpbc}
\sigma_{nn}({\bf k}) = \frac{d\theta}{\eta\sqrt{b^2-1}}
\frac{1+\mu^{-N}}{1-\mu^{-N}}, \qquad \mu = b + \sqrt{b^2-1} .
\end{equation}
In the region $1-\eta \ll 1$ and $a_0 k \ll 1$ 
this expression has the limiting forms
\begin{equation}\label{siglim}
\renewcommand{\arraystretch}{2.0}
\sigma_{nn}({\bf k}) \cong 
\left\{
\begin{array}{ll}
\displaystyle
\frac{2d\theta}{\eta N}\frac{1}{2d(1-\eta)+(a_0 k)^2},         
                 & L/r_c(k)=N\sqrt{2d(1-\eta)+(a_0 k)^2} \ll 1  \\
\displaystyle
\frac{d\theta}{\eta}\frac{1}{\sqrt{2d(1-\eta)+(a_0 k)^2}},
                 & L/r_c(k) \gg 1  
\end{array}
\right. 
\end{equation}
demonstrating the crossover from $d$- to $d'$-dimensional behaviour 
mentioned above. The second of these limiting expressions corresponds to 
the bulk and can also be obtained by the integration of the 
bulk CF $s_{\bf k}$ (\ref{cfhomo}) over the $d$th component 
of the wave vector.
For $N\gg 1$ and $d=3$ the integral in (\ref{sconstrfd}) 
with $\sigma_{nn}({\bf k})$ (\ref{sigpbc}) can be 
calculated analytically, and the result for 
$\theta_c$ reads (pbc)
\begin{equation}\label{tcpbc}
\theta_c^{-1} \cong W_3 + \frac{3}{\pi N} 
\ln\frac{1}{1-\exp[-N\sqrt{6(1-\eta)}]}.
\end{equation}
In the limit of extremely small anisotropies $1-\eta$ 
the transition temperature $\theta_c$ becomes logarithmically small:
\begin{equation}\label{tcpbcln}
\theta_c \cong \left( \frac{2\pi N}{3} \right)
\bigg/\ln{\frac{1}{6N^2(1-\eta)}} \ll 1,
\end{equation}
but this limit is very difficult to reach for $N\gg 1$. 
The minimal value of $1\!-\!\eta$ required to support $\theta_c\sim 1$ 
diminishes exponentially fast with the increase of $N$:
$1-\eta^* \sim N^{-2}\exp(-2\pi N/3)$. 
For $d=4$ the results have the form (pbc)
\begin{equation}\label{tc4dpbc}
\renewcommand{\arraystretch}{2.0}
\theta_c^{-1} \cong 
\left\{
\begin{array}{ll}
\displaystyle
W_4 + \frac{2}{3N^2},             & 1 \ll N^2 \ll 1/(1-\eta)  \\
\displaystyle
W_4 + \frac{4[2(1-\eta)]^{1/4}}{(\pi N)^{3/2}}
\exp{[-2N\sqrt{2(1-\eta)}]},           & N^2(1-\eta) \gg 1 .    
\end{array}
\right. 
\end{equation}
The first of these limiting expressions coinsides 
with that of Barber and Fisher \cite{barfis73}.

Now we proceed to the investigation of the more complicated case of a 
ferromagnetic film with free boundary conditions. Here the solution 
of the Dyson equation (\ref{dysoncffd}) with $n=n_0$ can be represented 
by the recurrence formula
\begin{equation}\label{sigrec}
\sigma_{nn} = \frac{2d\theta}{\eta}
\frac{1}{2b_n - \alpha_n - \alpha'_n}, \qquad
\alpha_{n+1} = \frac{1}{2b_n - \alpha_n}, \qquad
\alpha'_{n-1} = \frac{1}{2b_n - \alpha'_n}
\end{equation}
with the initial conditions
\begin{equation}\label{initcond}
\alpha_1 = \alpha'_N = 0, \qquad
\alpha_2 = 1/(2b_1), \qquad
\alpha'_{N-1} = 1/(2b_N) .
\end{equation}
Now all quantities $G_n$ entering $b_n$ (\ref{bn}) can be determined 
numerically as functions of $\theta$ from the $N$ constraint equations 
(\ref{sconstrfd}). Finally, $\theta_c$ can be found from the condition
$D_N=0$, where $D_N$ is the determinant of the linear system 
(\ref{mageqfd}).
The problem can be solved analytically in two limiting 
cases depending on the value of $N^2(1-\eta)$ (see (\ref{siglim})).
In the limit $N^2(1-\eta)\gg 1$ the system shows a $d$-dimensional 
(bulk) behavior in the whole range of ${\bf k}$, and in the main 
part of a sample all $\sigma_{nn}({\bf k})$ are equal to 
each other and determined by the value of $G$ far from the boundaries. 
Indeed, in this region the recurrence 
relations in (\ref{sigrec}) converge in the depth of the sample to 
$\alpha=\alpha'=b-\sqrt{b^2-1}$, which leads to the bulk expression for 
$\sigma_{nn}({\bf k})$ analogous to the second one in (\ref{siglim}).
The value of $G$ in the depth of the film is in our limit insensitive 
to its behaviour in the boundary regions $n, N\!-\!n \lesssim n_c \sim 1$
and can be found from the condition $D_N=0$ using $G_n=G=\const$.
To see that, one can simply cut the boundary regions and 
require $D_{N-n_c}=0$, which introduces corrections of the order 
$1/N\ll 1$. The calculation analogous to that in the MFA case yields
$G \cong 1 + (\pi/N)^2/(2d)$.  
After integration over ${\bf k}$ in (\ref{sconstrfd}) 
one arrives at the obvious expression $s_{nn}=\theta G P(\eta G)$
(see (\ref{cfhomo})),
and the value of $\theta_c$ determined from the 
condition $s_{nn}=1$ reads (fbc, $N^2(1-\eta)\gg 1$)
\begin{equation}\label{tcquasibulk}
\theta_c \cong \frac{1}{P(\eta)}
\left[
1 - \frac{1}{2d}
\left(\frac{\pi}{N}\right)^2 I(\eta)
\right] , \qquad
I(\eta) = 1 + \frac{\eta P'(\eta)}{P(\eta)} .
\end{equation}
For $d=3$ the limiting forms of $I(\eta)$ obtained from 
(\ref{plims}) are given by
\begin{equation}\label{ieta}
\renewcommand{\arraystretch}{1.5}
I(\eta) \cong
\left\{
\begin{array}{ll}
\displaystyle
1 + \eta^2/d ,       & \eta \ll 1
\\
\displaystyle
\frac{(3/2)^{3/2}}{\pi W_3}\frac{1}{\sqrt{1-\eta}} , 
& 1-\eta \ll 1 .
\end{array}
\right.
\end{equation}
One can see that (\ref{tcquasibulk}) 
generalizes the MFA result (\ref{tcmfa}), and 
for $1-\eta \ll 1$ corrections to $\theta_c$ due to the 
finite-size effects are much greater than in the MFA.

For $d\geq 5$ the derivative $P'(\eta)$ is finite for 
$\eta\to 1$ (see (\ref{plims})) and the results obtained above can be 
applied for all values of $\eta$. The reason for this is that the region 
of small wave vectors, $k\lesssim k_N \equiv a_0^{-1}/N$, where for 
$N^2(1-\eta)\lesssim 1$ the quasi-bulk expression for the CF 
$\sigma_{nn}({\bf k})$ becomes invalid (see (\ref{siglim})), is 
suppressed by the phase-volume factor in the integral 
(\ref{sconstrfd}). The marginal case 
is $d=4$, where for $1-\eta \lesssim (a_0 k_N)^2 \sim 1/N^2$
with the logarithmic accuracy it is sufficient to 
calculate the integral over the Brillouin zone down to $k_N$. As a 
result one gets (fbc)
\begin{equation}\label{tc4d}
\renewcommand{\arraystretch}{1.5}
\theta_c^{-1} \cong 
\left\{
\begin{array}{ll}
\displaystyle
W_4 + \frac{1}{N^2}\ln N + O\left( \frac{1}{N^2} \right),         
                 & 1 \ll N^2 \lesssim 1/(1-\eta)  \\
\displaystyle
W_4 + \frac{1}{2N^2}\ln\frac{c'_4}{1-\eta} ,
                 & N^2(1-\eta) \lesssim 1 ,    
\end{array}
\right. 
\end{equation}
(cf. (\ref{tc4dpbc})). An asymptotic dependence of the type 
$\ln(N)/N^2$ in the isotropic limit with a coefficient 
close to unity was obtained numerically for the 
spherical model with the layer-constraint in \cite{cosmazmih76}.  
To the contrast, for the standard spherical model \cite{berkac52} 
with fbc $\theta_c^{-1} \cong  W_4 + a/N$ with $a<0$ \cite{barfis73}.

For a 3-dimensional ferromagnetic film in the limit 
$N^2(1-\eta)\ll 1$ the leading 
correction to the ${\bf k}$-integral (\ref{sconstrfd}) and hence to 
$\theta_c$ comes from the longwavelength region 
$k\lesssim k_N = a_0^{-1}/N$, where 
$\sigma_{nn}({\bf k})$ behaves 2-dimensionally. The form of 
$\sigma_{nn}({\bf k})$ in this region can be determined from the general 
formula (\ref{sigrec}). Beyond the narrow boundary regions the 
quantities $\bar b_n\equiv b_n(\eta=1,{\bf k}=0)$, etc., satisfy
$\bar b_n \cong \bar \alpha_n \cong \bar \alpha'_n \cong 1$, and the 
values of $\alpha_n$ and $\alpha'_n$ can be found from the recurrence 
relations (\ref{sigrec}) with the help of the expansion with respect to 
small $1-\eta$ and $(a_0 k)^2$. As a result on gets the same expression 
(\ref{siglim}) in the same wave vector range $k\lesssim k_N$. This region 
yields the contribution of the order $(1/N)\ln[1/(N^2(1-\eta))]$ into 
$s_{nn}$ (\ref{sconstrfd}). The contribution of the region
$k\gtrsim k_N$ into the correction to $s_{nn}$ can be estimated in the 
following way. For $1-\eta \ll 1$ the finite-size correction 
described by (\ref{tcquasibulk}) and (\ref{ieta}) comes from the region
of small wave vectors $a_0 k\sim \sqrt{1-\eta}$. In our case, however, 
$a_0 k_N\gg \sqrt{1-\eta}$, and the corresponding contribution is 
reduced to the value of the order $1/N$. With the logarithmic accuracy 
the latter can be neglected in comparison to that of the 2-dimensional 
region $k\lesssim k_N$. The final result for $\theta_c$ of the 
3-dimensional model with free boundary conditions 
can be written as (fbc, $N^2(1-\eta)\ll 1$)
\begin{equation}\label{tcfbc}
\theta_c^{-1} \cong W_3 + \frac{3}{2\pi N} 
\ln\frac{1}{N^2(1-\eta)} + O\left( \frac{1}{N} \right) .
\end{equation}
The similarity of this result with (\ref{tcpbc}) is not surprising 
since in the relevant region, $k\lesssim k_N$, where
$r_c(k)\gtrsim L$, all N layers are strongly correlated with each other 
and the type of boundary conditions plays no role in the 
leading approximation.

The author thanks Hartwig Schmidt for valuable discussions.
The financial support of Deutsche Forschungsgemeinschaft 
under contract Schm 398/5-1 is greatfully acknowledged.



\end{document}